\documentclass[aps,prl,twocolumn,showpacs,amsmath,amssymb]{revtex4}
\usepackage{bm}
\usepackage{graphicx}
\usepackage{times}
\usepackage{epstopdf}
\begin{document}

\makeatletter
\newcommand*{\balancecolsandclearpage}{%
  \close@column@grid
  \clearpage
  \twocolumngrid
}
\makeatother

\renewcommand{\section}[1]{{\par\it #1.---}\ignorespaces}
\newcommand{\did}{$d_1 \! + \! id_2$}

\title{Edge properties and Majorana fermions in the proposed chiral $d$-wave superconducting state of doped graphene}
\author{Annica M. Black-Schaffer}
 \affiliation{Department of Physics and Astronomy, Uppsala University, Box 516, S-751 20 Uppsala, Sweden}
\date{\today}

\begin{abstract}
We investigate the effect of edges on the intrinsic $d$-wave superconducting state in graphene doped close to the van Hove singularity. While the bulk is in a chiral $d_{x^2-y^2}+id_{xy}$ state, the order parameter at any edge is enhanced and has  $d_{x^2-y^2}$-symmetry, with a decay length strongly increasing with weakening superconductivity. 
No graphene edge is pair breaking for the $d_{x^2-y^2}$ state and there are no localized zero-energy edge states.
We find two chiral edge modes which carry a spontaneous, but not quantized, quasiparticle current related to the zero-energy momentum. Moreover, for realistic values of the Rashba spin-orbit coupling, a Majorana fermion appears at the edge when tuning a Zeeman field.
\end{abstract}
\pacs{74.20.Rp, 74.70.Wz, 74.20.Mn, 73.20.At, 71.10.Pm}
\maketitle

%
Graphene, a single layer of carbon, has generated immense interest ever since its experimental discovery \cite{Novoselov04}. Lately, experimental advances in doping methods \cite{McChesney10,Efetov10} have allowed the electron density to approach the van Hove singularities (VHSs) at 25\% hole or electron doping. The logarithmically diverging density of states (DOS) at the VHS can allow non-trivial ordered ground-states to emerge due to strongly enhanced effects of interactions. Very recently, both perturbative renormalization group (RG) \cite{Nandkishore11} and functional RG calculations \cite{Kiesel11,Wang11} have shown that a chiral spin-singlet $d_{x^2-y^2}+id_{xy}$ (\did) superconducting state likely emerges from electron-electron interactions in graphene doped to the vicinity of the VHS. This is in agreement with earlier studies of strong interactions on the honeycomb lattice near half-filling \cite{Black-Schaffer07,Honerkamp08,Pathak10,Ma11}.

Rather unique to the honeycomb lattice is the degeneracy of the two $d$-wave pairing channels \cite{Black-Schaffer07,Gonzalez08}. Below the superconducting transition temperature ($T_c$), this degeneracy results in the time-reversal symmetry breaking \did\ state \cite{Black-Schaffer07, Nandkishore11}.
However, any imperfections, and most notably edges, might destroy this degeneracy and generate a local superconducting state different from that in the bulk. At the same time, many of the exotic features proposed for a \did\ superconductor, such as spontaneous \cite{Volovik97,Fogelstrom97}, or even quantized \cite{Laughlin98}, edge currents and quantized spin- and thermal Hall effects \cite{Senthil99, Horovitz03}, are intimately linked to its edge states. 
In order to determine the properties of \did\ superconducting graphene, it is therefore imperative to understand the effect of edges on the superconducting state.

In this Letter we establish the edge properties of \did\ superconducting graphene doped to the vicinity of the VHS. We show that, while the bulk is in a \did\ state, any edge will be in a pure, and enhanced, $d_{1}$-wave state. Due to a very long decay length of the edge $d_1$ state, the edges influence even the properties of macroscopic graphene samples. 
We find two well-localized chiral edge modes which carry a spontaneous, but not quantized, edge current. 
Furthermore, we show that by including a realistic Rashba spin-orbit coupling, graphene can be tuned, using a Zeeman field, to host a Majorana fermion at the edge.
These results establish the exotic properties of the chiral \did\ superconducting state in doped graphene, which if experimentally realized, will provide an exemplary playground for topological superconductivity. 
Furthermore, these results are also very important for any experimental scheme aimed at detecting the \did\ state in graphene, as such scheme will likely be based on the distinctive properties of the edge.

\begin{figure}[htb]
\includegraphics[scale = 0.8]{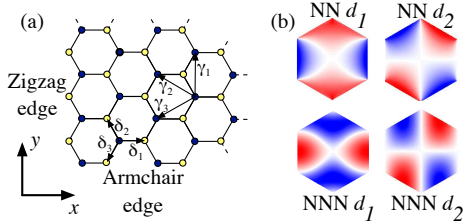}
\caption{\label{fig:gap} (Color online) (a) Graphene with NN bonds $\delta_\alpha$, NNN bonds $\gamma_\alpha$, zigzag and armchair edges indicated. 
(b) Different $d$-wave superconducting order parameters for NN and NNN pairing with negative (blue) and positive (red) sign.
}
\end{figure}

We approximate the $\pi$-band structure of graphene as:
\begin{align}
\label{eq:H0}
H_0 = -t  \sum_{\langle i,j\rangle,\sigma}  c^\dagger_{i\sigma}c_{j\sigma} + \mu \sum_i c^\dagger_{i\sigma} c_{i\sigma},
\end{align}
where $t = 2.5$~eV is the nearest neighbor (NN) hopping amplitude and $c_{i\sigma}$ is the annihilation operator on site $i$ with spin $\sigma$. The chemical potential is $\mu$ and the VHS appears at $\mu = \pm t$, where the Fermi surface transitions from being centered around $K$, $K'$ to $\Gamma$.
 We study two different models for superconducting pairing from repulsive electron-electron interactions:
\begin{align}
\label{eq:HDelta}
H_\Delta = \sum_{i,\alpha} \Delta_{\alpha}(i)(c^\dagger_{i\uparrow}c^\dagger_{i+a_\alpha\downarrow} - c^\dagger_{i\downarrow}c^\dagger_{i+a_\alpha\uparrow}) + {\rm H.c.}.
\end{align}
In the limit of very strong on-site Coulomb repulsion (mean-field) pairing appears on NN bonds such that $a_\alpha = \delta_\alpha$ ($\alpha = 1,2,3$) \cite{Black-Schaffer07}, whereas a moderate on-site repulsion gives rise to pairing on next-nearest-neighbor (NNN) bonds with $a_\alpha = \gamma_\alpha$ \cite{Kiesel11}, see Fig.~\ref{fig:gap}(a).
The high electron density near the VHS efficiently screen long-range electron-electron interactions, and we also show that our results are largely independent on the choice of $a$. 
In mean-field theory the order parameter can be calculated from the condition $\Delta_{\alpha}(i) = -J\langle c_{i\downarrow} c_{i+a_\alpha \uparrow} - c_{i\uparrow} c_{i+a_\alpha \downarrow} \rangle$. Here $J$ is the effective (constant) pairing potential arising from the electron-electron interactions and residing on NN bonds for $a = \delta$ and on NNN bonds for $a = \gamma$. Using this condition for $\Delta$, the Hamiltonian $H = H_0 +H_\Delta$ can be solved self-consistently within the Bogoliubov-de Gennes formalism \cite{Black-Schaffer08,SMnote}. 
The favored bulk solution of $\Delta_\alpha$ belongs to the two-dimensional $E_{2}$ irreducible representation of the $C_{6v}$ lattice point group. This representation can be expressed in the basis ${\hat a}_{d_1} = (1,-\frac{1}{2},-\frac{1}{2})$, which has $d_1 = d_{x^2-y^2}$ symmetry when $H_0$ is diagonal, and ${\hat a}_{d_2} = (0,\frac{\sqrt{3}}{2},-\frac{\sqrt{3}}{2})$ which has $d_2 = d_{xy}$ symmetry, see Fig.~\ref{fig:gap}(b). In the translational invariant bulk, these two solutions have the same $T_c$, but below $T_c$ the complex combination \did\ has the lowest free energy \cite{Black-Schaffer07,Nandkishore11}. There is also an $s$-wave solution, ${\hat a}_s = (1,1,1)$, but it only appears subdominantly and at very strong pairing.
%
%

In order to quantify the edge effects we study thick ribbons with both zigzag and armchair edges. We assume smooth edges and Fourier transform in the direction parallel to the edge. Due to computational limitations we need $J\gtrsim 0.5t$ in order to reach bulk conditions inside the slab. This gives rather large $\Delta_\alpha$, but by studying the $J$-dependence we can nonetheless draw conclusions for the experimentally relevant low-$J$ regime.

%
\section{Superconducting state at the edge}
\begin{figure}[htb]
\includegraphics[scale = 0.9]{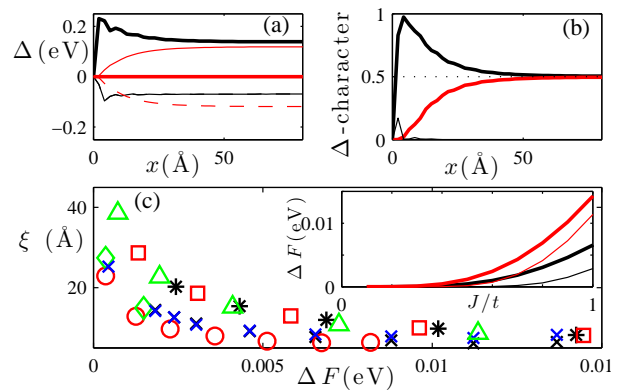}
\caption{\label{fig:DJa} (Color online) (a) Order parameter profile for the zigzag edge for NN $J = 0.75t$ at the VHS with real (black) and imaginary (red) part for $\Delta_1$ (thick), $\Delta_2$ (thin), and $\Delta_3$ (dashed) [black dashed line is hidden behind black solid line since ${\rm Re}(\Delta_2) \approxeq {\rm Re}(\Delta_3)$].
(b) Character of the order parameter in (a): $d_1$ (thick black), $d_2$ (thick red), and $s$ (thin black). Dotted line marks the bulk value.
(c) Decay length $\xi$ of the $d_1$-character as function of $\Delta F$ for different doping levels, edges, and superconducting pairing: NN pairing, zigzag edge, and $\mu = t$ (black $\times$), $\mu = 0.8t$ (red $\circ$), $\mu = 1.2t$ (green $\diamond$) or armchair edge and $\mu = t$ (blue $\times$), NNN pairing, zigzag edge, and $\mu = t$ (black $\star$), $\mu = 0.8t$ (red $\square$), $\mu = 1.2t$ (green $\triangle$) [blue, $\times$ symbols is often completely overlaying black, $\times$ symbols since no notable difference is found between zigzag and armchair edges]. Inset shows $\Delta F$ as function of the pairing potential for NN pairing (black) and NNN pairing (red) for $\mu = t$ (thick) and $\mu = 1.2t$ (thin). $\mu < t$ has a $\Delta F$ curve similar to $\mu > t$.
}
\end{figure}

In the bulk, the \did\ state has a free energy $\Delta F$ lower than the $d_{1,2}$ states, which varies strongly with both doping and pairing potential, see inset in Fig.~\ref{fig:DJa}(c). However, sample edges break the translational invariance and a qualitatively different solution emerges. Figure \ref{fig:DJa}(a) shows how the zigzag edge completely suppresses the imaginary part of $\Delta_\alpha$, while at the same time enhancing the magnitude.
This suppression leads to a pure $d_1$ solution at the edge, an effect we quantify in Fig.~\ref{fig:DJa}(b) by plotting the $d_1$-character $|\frac{\sqrt{2}}{\sqrt{3}}(\Delta_1-\frac{1}{2}(\Delta_2+\Delta_3))|^2$. The edge behavior can be understood by noting that bonds $\delta_2$ and $\delta_3$ ($\gamma_2$ and $\gamma_3$) are equivalent for both armchair and zigzag edges \cite{Black-Schaffer09} and, therefore, the $d_1$-wave state is heavily favored at both type of edges. Since the edge is of the zigzag type for edges with  $30^\circ$ and $90^\circ$ angles off the $x$-axis and of the armchair type for $0^\circ$ and $60^\circ$ angles, we conclude that any edge should host $d$-wave order with nodes angled $45^\circ$ from the edge direction.  
%
%
In order to quantify the spatial extent of this edge effect, we calculate a decay length $\xi$ by fitting the $d_1$-character profile to the functional form $(Ce^{-x/\xi} + 0.5)$ with $C \approx 0.5$. 
As seen in Fig.~\ref{fig:DJa}(c), $\xi$ varies strongly with $\Delta F$, but very little with edge type and doping level. 
Furthermore, the increase in $\xi$ for NNN pairing compared to NN pairing suggests that the edge will be even more important in models with longer ranged Coulomb repulsion.
The strongly increasing $\xi$ with decreasing $\Delta F$ has far-reaching consequences for graphene. For example, $J = 0.5t$ and doping at the VHS gives $\xi \approx 25$~\AA\ for NN pairing. With an expected much weaker superconducting pairing in real graphene, the edge will not only modify the properties of the superconducting state in graphene nanoribbons, but also in macroscopically sized graphene samples.
We have verified that both the \did\ state itself and edge effects described here are stable in the presence of random disorder \cite{SMnote}.

%
\section{Chiral edge states}
\begin{figure}[htb]
\includegraphics[scale = 0.9]{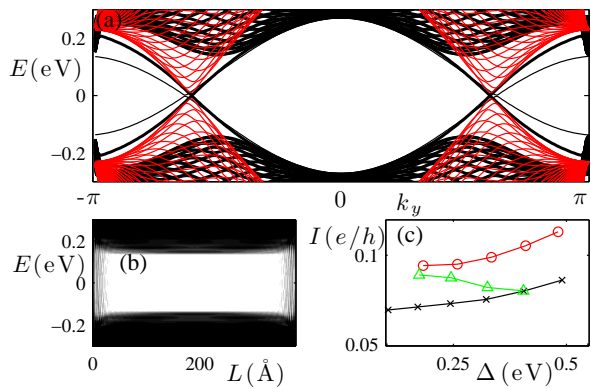}
\caption{\label{fig:Ek} (Color online) (a) Band structure for a zigzag edge slab with NN $J = 0.75t$, $\mu = t$, and self-consistent $\Delta$ (thick black), constant \did\ state corresponding to the bulk state (thin black), and constant $d_1$ state corresponding in amplitude to the $d_1$ state at the surface.
(b) LDOS across the ribbon for the self-consistent solution in (a) interpolating between  0.2 (black) to 0 (white) states/eV/unit cell, showing a bulk gap of 0.18 eV and gapless edge states.
(c) Quasiparticle edge current in units of $e/h$ as function of superconducting bulk order parameter $\Delta(1,e^{2\pi i /3},e^{4\pi i/3}$) for zigzag edge with $\mu = t$ (black $\times$), $\mu = 0.8t$ (red $\circ$), and armchair edge with $\mu = t$ (green $\triangle$).
}
\end{figure}

Any \did\ state, even with one subdominant part, violates both time-reversal and parity symmetry and has been shown to host two chiral edge states \cite{Volovik97, Laughlin98, Senthil99}. The topological invariant guaranteeing the existence of these two chiral edge modes also causes quantized spin- and thermal-Hall responses \cite{Senthil99,Horovitz03}.
Figure~\ref{fig:Ek}(a) shows the band structure for a zigzag slab. The self-consistent solution (thick black) gives two Dirac cones located at $\pm k_0$, where bands with same velocities reside on the same surface, thus yielding two co-propagating chiral surface states per edge. 
The band structure for the constant (non-self-consistent) bulk \did\ state also has two Dirac cones (thin black), but shifted away from $k_0$. The shift is directly related to the $d_1$ state at the edge. The $d_1$ state has no surface states on the zigzag edge, only bulk nodal quasiparticles, where the nodes for a  $d_1$ order parameter with amplitude equal to that on the edge are located at $\pm k_0$ (thin red).
%
The similarity between the \did\ and $d_1$ edge band structures thus makes for only modest effects of the edge on the self-consistent band structure. It also results in the chiral edge modes being well localized to the edge, as seen in the local density of states (LDOS) plot in Fig.~\ref{fig:Ek}(b). The constant edge LDOS is a consequence of the one-dimensional Dirac spectrum.
We note especially that no $d$-wave superconducting graphene edge will display a zero-bias conductance peak due to zero-energy surface states, in contrast to the cuprate superconductors \cite{Fogelstrom97}. 
Such a peak is only present when the order parameter for incidence angle $\theta$ on the edge has a different sign from when the angle is $\pi-\theta$. This only happens for the $d_2$-solution on both the zigzag and armchair edge.

The breaking of time-reversal symmetry gives rise to spontaneous edge currents carried by the chiral edge modes \cite{Fogelstrom97,Volovik97,Laughlin98,Horovitz03}. 
By combining the charge continuity equation with the Heisenberg equation for the particle density \cite{Black-Schaffer08}, we calculate in Fig.~\ref{fig:Ek}(c) the quasiparticle edge current as function of of the bulk order parameter $\Delta(1,e^{2\pi i /3},e^{4\pi i/3})$.
We find no evidence for a quantized boundary current equal to $2e \Delta/h$, as previously suggested \cite{Laughlin98}. In fact, we find a non-linear relationship between current and $\Delta$, a strong variation with doping level, and, most importantly, the armchair current even {\it decreases} when $\Delta$ increases. The last result can be understood by studying the zero-energy crossing $\pm k_0$ of the chiral edge modes. For the zigzag edge $k_0$ increases with increasing $\Delta$, whereas for the armchair edge $k_0$ decreases. In general, we find that changes in current are proportional to $\delta k_0^\beta$ with $\beta \approx 1-2$. This, at least, partially agree with earlier results reporting a $\beta = 2$ dependence \cite{Volovik97}. Finite $k$-point sampling and neglecting the screening supercurrents could potentially explain the discrepancy.

%
\section{Majorana mode}
Heavy doping of graphene, by either ad-atom deposition \cite{McChesney10} or gating \cite{Efetov10}, breaks the $z \rightarrow -z$ mirror symmetry and introduces a Rashba spin-orbit coupling \cite{Kane05b}
\begin{align}
\label{eq:Hl}
H_{\lambda} = i\lambda_R \! \! \sum_{\langle i,j\rangle,\sigma,\sigma'} \hat{z}\cdot ({\bf s}_{\sigma,\sigma'} \times {\bf \hat{d}}_{ij}) c^\dagger_{i\sigma} c_{j \sigma'},
\end{align}
where ${\bf \hat{d}}_{ij}$ is the unit vector from site $j$ to $i$.
Superconducting two-dimensional systems with Rashba spin-orbit coupling and magnetic field have recently attracted much attention due to the possibility of creating Majorana fermions at vortex cores or edges \cite{Sau10, Alicea10, Sato10}. At edges the Majorana fermion appears as a single mode crossing the bulk gap. This should be contrasted with the behavior found above, where the edge instead hosts two modes. 
We will here show that a Majorana mode is created in $d$-wave superconducting doped graphene in the presence a moderate Zeeman field: $H_h = -h_z\sum_{i}(c^\dagger_{i\uparrow} c_{i \uparrow}-c^\dagger_{i\downarrow} c_{i\downarrow})$.
Due to spin-mixing in $H_{\lambda}$, the basis vector $X^\dagger = (c^\dagger_{i\uparrow} c^\dagger_{i\downarrow} c_{i\uparrow}c _{i \downarrow})$ has to be used when expressing the Hamiltonian $H_{\rm ext} = H_0+H_\Delta + H_\lambda + H_h$ in matrix form: $H_{\rm ext} = \frac{1}{2}X^\dagger \mathcal{H_{\rm ext}} X$. This results in a doubling of the number of eigenstates compared to the physical band structure. This doubling is necessary for the appearance of the Majorana fermion, since a regular fermion consists of two Majorana fermions. 

\begin{figure}[htb]
\includegraphics[scale = 0.95]{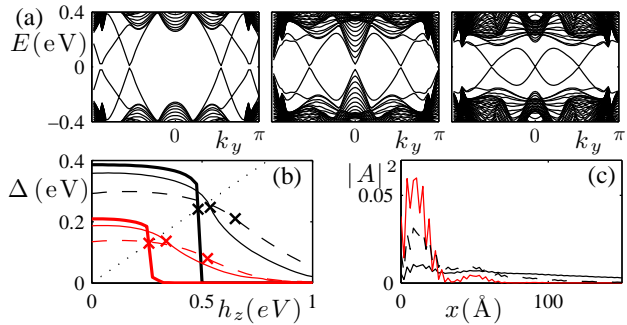}
\caption{\label{fig:Maj} (Color online) (a) Eigenvalue spectrum for a zigzag slab with NN $J = 1.2t, \mu = t, \lambda_R = 0.2t$, and $h_z = 0.4, 0.535$ and $0.6$~eV (left to right), with $h_c = 0.535$~eV. Small gaps in the surface states are due to limited $k$-point sampling.
(b) Self-consistent $\Delta$ as function of $h_z$ for $J = 1.2t$ (black), $0.9t$ (red) for $\lambda_R = 0.05t$ (thick), $0.2t$ (thin), and $0.3t$ (dashed). Dotted line mark the $h_c = 2\Delta$ one-band model phase transition. Crosses mark the numerical phase transition. 
(c) Eigenvalue amplitude squared for the Majorana mode in (a) for $h_z = 0.54$~eV (black), $0.56$ (dashed), and $0.6$ (red).
}
\end{figure}

A change in the number of edge modes marks a topological phase transition which, in general, can only occur when the bulk energy gap closes. We therefore start by identify the conditions for bulk zero energy solutions of $H_{\rm ext}$. 
Close to the VHS we can, to a first approximation, use only the partially occupied $\pi$-band for small $\Delta,\lambda_R,$ and $h_z$. A straightforward calculation \cite{Sato10} for this one-band Hamiltonian gives the following bulk-gap closing conditions at $\mu \sim t$:
\begin{align}
\label{eq:Eg}
(\mu - t|\epsilon_k|)^2 + \Delta_k^2 = h_z^2 + \lambda_R^2|{\mathcal L}_k|^2, \ \ \ |\Delta_k||\lambda_R \mathcal{L}_k| = 0,
\end{align}
where $\epsilon_k = \sum_{\alpha}e^{ik \delta_\alpha}$ is the band structure, $\varphi_k = {\rm arg}(\epsilon_k)$,  $\Delta_k = -\sum_\alpha \Delta_\alpha \cos(k\delta_\alpha - \varphi_k)$ is the $k$-dependent intraband superconducting order for NN pairing \cite{Black-Schaffer07}, and $\mathcal{L}_k = {\rm Im}[e^{-i\varphi_k}(-\frac{\sqrt{3}}{2}e^{ik\delta_2} + \frac{\sqrt{3}}{2}e^{ik\delta_3},e^{ik\delta_1}-\frac{1}{2}e^{ik\delta_2}-\frac{1}{2}e^{ik\delta_3},0)]$ is the spin-orbit interaction when expressed in the form $H_\lambda = \sum_{k\sigma \sigma'} \lambda_R\mathcal{L}_k\cdot {\bf s}_{\sigma \sigma'} c^\dagger_{k \sigma} c_{k \sigma'}$ for the one-band model.
Equations~(\ref{eq:Eg}) are met at $\Gamma, K$, and $M$ in the Brillouin zone, where they produce the conditions $(\mu - 3t)^2 = h_z^2$, $\mu^2 = h_z^2$, and $(\mu - t)^2 + \Delta_k^2(M) = h_z^2$, respectively. At $\mu \sim t$ only the last condition is satisfied for small $h_z$, which is necessary for superconductivity to survive.
We find $\Delta_k(M) = 2\Delta$ for the $\Delta(1,e^{2\pi i /3},e^{4\pi i/3})$ order parameter and, thus, at the VHS there is a topological phase transition at $h_c = 2\Delta$. Figure~\ref{fig:Maj}(a) shows how the eigenvalue spectrum of a superconducting zigzag slab at the VHS develops when $h_z$ is swept past $h_c$. 
At finite $\lambda_R$ and/or $h_z$ the chiral modes in Fig.~\ref{fig:Ek}(a) split with one mode moving towards $k_y = 0$ and the other one towards the zone boundary at $k_y = \pi$, see left-most figure in Fig.~\ref{fig:Maj}(a). At $h_c$ (center figure) the bulk gap closes at both $k_y = 0,\pi$. The closure at $k_y = \pi$ annihilates the outer chiral modes whereas the closure at $k_y=0$ leaves a new Dirac cone crossing the bulk band gap with the two modes belonging to different edges. Thus, at $h_z>h_c$ we are left with three modes per edge crossing the bulk gap. The odd number establishes the existence of a Majorana mode alongside the two remnant chiral modes.
%
Figure~\ref{fig:Maj}(b) shows how $\Delta$ develops in the presence of an applied Zeeman field $h_z$, with $\times$-symbols marking the phase transition into the phase with a Majorana fermion. The dotted line marks the one-band result $h_c = 2\Delta$, which is a good approximation for small $\lambda_R$. In this small $\lambda_R$-regime there is a very pronounced drop in $\Delta$ at the phase transition with only a small remnant superconducting state in the Majorana phase at $h_z>h_c$, which results in a poorly resolved Majorana mode. 
Larger $\lambda_R$ gives a stronger superconducting state in the Majorana phase. However, for $\lambda_R > 0.2t$ we find $h_c>2\Delta$, and the superconducting state is again very weak beyond the phase transition. 
We thus conclude that, in order to create a Majorana fermion at the edge of $d$-wave superconducting graphene doped very close to the VHS, a small to moderate Rashba spin-orbit coupling, $\lambda_R\sim 0.2t$, and a Zeeman field of the order of $2\Delta$ is needed. 
With reported tunability with electric field \cite{Min06}, as well as impurity-induced Rashba spin-orbit coupling \cite{CastroNeto09b}, $\lambda_R \sim 0.2t$ is likely within experimental reach in heavily doped graphene. The Zeeman field can be generated by proximity to a ferromagnetic insulator, whereas if applying an external magnetic field, orbital effects also needs to be taken into account.
 Finally, in Fig.~\ref{fig:Maj}(c) we plot the spatial profile of the Majorana mode amplitude just beyond $h_c$. Due to the larger $\Delta$ at the edge, the bulk enters the Majorana-supporting topological phase before the edge. Therefore, the Majorana mode does not appear at the edge but is spread throughout the sample for $h_z\gtrsim h_c$. Not until $h_z>2\Delta({\rm edge})$ does the Majorana mode appear as a pure edge excitation.

In summary, we have shown that the \did\ superconducting state in heavily doped graphene is in a pure $d_1$ state on any edge. 
The $d_1$ edge state significantly modifies the superconducting state even in macroscopic graphene samples due to a long decay length.
Moreover, \did\ superconducting graphene hosts two well-localized chiral edge modes, which carry a non-quantized spontaneous quasiparticle current. A Majorana mode can also be created at the edge by tuning a moderate Zeeman field. 
These results establish the properties of the \did\ state in graphene, and will be important for any experimental detection of this state.

\begin{acknowledgments}
The author thanks A.~V.~Balatsky, M.~Fogelstr\"om, and T.~H.~Hansson for discussions and the Swedish research council (VR) for support. 
\end{acknowledgments}


\balancecolsandclearpage

\onecolumngrid
\begin{center} {\large \bf Supplementary material}\end{center}
In this supplementary material we provide: (1) a detailed, largely self-contained, description of the method underlying our results, and (2) numerical data showing the relative robustness of the bulk \did\ superconducting state and its edge properties in the presence of disorder.

\section{Method}
\label{sec:method}
As described in the main text, we use the Hamiltonian $H = H_0 + H_\Delta$, where 
\begin{align}
\label{eq:Hall}
H_0 & = -t  \sum_{\langle i,j\rangle,\sigma}  c^\dagger_{i\sigma}c_{j\sigma} + \mu \sum_i c^\dagger_{i\sigma} c_{i\sigma}, \\
H_\Delta & = \sum_{i,\alpha} \Delta_{\alpha}(i)(c^\dagger_{i\uparrow}c^\dagger_{i+a_\alpha\downarrow} - c^\dagger_{i\downarrow}c^\dagger_{i+a_\alpha\uparrow}) + {\rm H.c.}.
\end{align}
Here $c_{i\sigma}$ is the annihilation operator on site $i$ of the honeycomb lattice with spin $\sigma$, $t =2.5$~eV is the nearest-neighbor (NN) hopping amplitude, and $\mu$ is the chemical potential, where $\mu = \pm t$ corresponds to the van Hove singularities (VHSs) ($H$ is particle-hole symmetric so hole and electron doping give the same result).
Furthermore, the superconducting order parameter $\Delta_\alpha$ resides on NN bonds when $a_\alpha = \delta_\alpha$ and on next-nearest neighbor (NNN) bonds when $a_\alpha = \gamma_\alpha$, where $\alpha = 1,2,3$ labels the three inequivalent bond directions, see Fig.~1(a) in the main text.
Within mean-field theory, $\Delta_\alpha$ is defined by the self-consistent condition:
\begin{align}
\label{eq:Deltaself}
\Delta_{\alpha}(i) = -J\langle c_{i\downarrow} c_{i+a_\alpha \uparrow} - c_{i\uparrow} c_{i+a_\alpha \downarrow} \rangle, 
\end{align}
where $J$ is the effective pairing potential on NN bonds for $a = \delta$ and on NNN bonds for $a = \gamma$. $J$ is a consequence of the local repulsive Coulomb interaction, which in the limit of very strong on-site repulsion results in pairing on NN bonds \cite{Black-Schaffer07} whereas a moderate on-site repulsion results in NNN bond pairing \cite{Kiesel11}.

We can solve $H$ within the Bogoliubov-de Gennes formalism by writing 
\begin{align}
\label{eq:HBdG}
H = X^\dagger \mathcal{H} X \ \ {\rm  with } \ \ X^\dagger = (c_{i,\uparrow}^\dagger, c_{i,\downarrow}),
\end{align}
and diagonalizing the matrix $\mathcal{H}$ to find all eigenvalues $E^\nu$ and eigenvectors $V^\nu$, where $\nu = 1, ... ,2N$ for $N$ sites. We can then define new operators $Y^\dagger = (\gamma^\dagger_\nu)$ using $X = \mathcal{V} Y$ with the columns of $\mathcal{V}$ given by the eigenvectors $V^\nu$, such that the Hamiltonian $H$ is diagonal in these new operators: $H = \sum_\nu E^\nu \gamma_\nu^\dagger \gamma_\nu$.
A self-consistent solution scheme start with first guessing the value of $\Delta_\alpha$, diagonalizing $\mathcal{H}$ for this value, using the self-consistent condition Eq.~(\ref{eq:Deltaself}) to recalculate $\Delta_\alpha$ from the eigenvalues and eigenvectors, and then reiterate these steps until the order parameter $\Delta_\alpha$ within two subsequent steps changes less than a small predetermined convergence limit.
Using the self-consistent value for $\Delta_\alpha$ any electronic property of the system can be calculated in using the eigenbasis. 
For example, the local density of states (LDOS) can be calculated as
\begin{align}
\label{eq:LDOS}
D_i(E) = \sum_\nu |V^\nu_i|^2 \delta(E-E^\nu) + |V^\nu_{N+i}|^2 \delta(E+E^\nu),
\end{align}
where the first part is the spin-up contribution and the second part the spin-down contribution. Numerically, we use a small Gaussian broadening for the $\delta$-functions.
We are also interested in the quasiparticle current, which can be calculated using the continuity equation for the charge current density ${\bf J}$:
\begin{align}
\label{eq:conteq}
{\bf \nabla} \cdot {\bf J} + \frac{\partial \rho}{\partial t} = 0
\end{align}
together with the Heisenberg equation for the particle density per unit cell $n_i$:
\begin{align}
\label{eq:Heisenberg}
\frac{d n_i}{d t} = \frac{i}{\hbar} [H, n_i],
\end{align}
where $\rho  = e\langle n \rangle$ \cite{Covaci06, Black-Schaffer08}. The quantum average of the commutator in Eq.~(\ref{eq:Heisenberg}) can easily be shown to only contain $H_0$ for a self-consistent solution of $\Delta_\alpha$. The total quasiparticle edge current is then simply $I = \sum {\bf J}_{||}$, where the summation is over all unit cells at the edge with a finite ${\bf J}$ parallel to the edge.

The above formalism can be applied to any structure on the honeycomb lattice. 
In order to investigate edge properties, we study $H$ on thick graphene ribbons having either zigzag or armchair edges. We make sure that the ribbons are always thick enough to guarantee bulk conditions in the interior. For simplicity, we assume smooth edges so we can Fourier transform in the direction along the edge, which introduces a $k$-point index, while reducing the site index $i$ to only enumerate sites perpendicular to the edge, i.e.~$i$ now only measures the distance to the edge.

We also study the influence of a finite Rashba spin-orbit coupling:
\begin{align}
\label{eq:Hlsupp}
H_{\lambda} = i\lambda_R \! \! \sum_{\langle i,j\rangle,\sigma,\sigma'} \hat{z}\cdot ({\bf s}_{\sigma,\sigma'} \times {\bf \hat{d}}_{ij}) c^\dagger_{i\sigma} c_{j \sigma'},
\end{align}
where ${\bf \hat{d}}_{ij}$ is the unit vector from site $j$ to $i$, in combination with a Zeeman field:
\begin{align}
\label{eq:Hz}
H_h = -h_z\sum_{i}(c^\dagger_{i\uparrow} c_{i \uparrow}-c^\dagger_{i\downarrow} c_{i\downarrow}).
\end{align}
The spin-mixing in the Rashba term now requires us to write $H_{\rm ext} = H_0+H_\Delta + H_\lambda + H_h$ as
\begin{align}
\label{eq:HBdG2}
H_{\rm ext} = \frac{1}{2}X^\dagger \mathcal{H}_{\rm ext} X \ \ {\rm  with } \ \ X^\dagger = (c^\dagger_{i\uparrow} c^\dagger_{i\downarrow} c_{i\uparrow}c _{i \downarrow}),
\end{align}
i.e.~we need to double the number of eigenstates compared to the physical band structure. This is expected since a Majorana fermion is essentially half a fermion.
By applying the same self-consistent procedure described above to $H_{\rm ext}$, we can solve for the superconducting order parameter $\Delta$, calculate all physical observables such as the LDOS, and also calculate the eigenvalue spectrum, which contains the Majorana mode for a large enough field $h_z$.
The Majorana mode appears when the eigenvalue spectrum develops from having an even number of edge modes to an odd number. Such a change in the number of edge modes is in general always associated with the closing of the bulk gap. We can analytically extract the approximate bulk gap closing condition from an effective one-band model. The kinetic Hamiltonian $H_0$ is diagonalized in the bulk by changing the basis from the site-operators $\{c_A, c_B\}$ on the two inequivalent sites $A$ and $B$, to the band operators $\{a,b\}$ through:
%
\begin{align}
\label{eq:banddiag}
\left( \begin{array}{c}
c_{A{k}\sigma}  \\ c_{B{k}\sigma}
\end{array} \right) = \frac{1}{\sqrt{2}}
\left( \begin{array}{c}
a_{{k}\sigma} + b_{{k}\sigma} \\ e^{-i\varphi_{ k}}(a_{{k}\sigma} - b_{{k}\sigma})
\end{array} \right).
\end{align}
Here $a_{{k}\sigma}^\dagger$ creates an electron in the lower $\pi$-band and $b_{{k}\sigma}^\dagger$ creates an electron in the upper $\pi$-band, such that 
%
\begin{align}
\label{eq:H0k}
H_{0}  & = \sum_{{k}\sigma} \left[ (\mu - t\epsilon_{k})a_{{k}\sigma}^\dagger a_{{k}\sigma} + (\mu + t\epsilon_{ k})b_{{ k}\sigma}^\dagger b_{{k}\sigma}  \right],
\end{align}
where the $k$-dependence of the $\pi$-bands is given by $\epsilon_{ k} =|\sum_{\alpha}e^{i{ k \cdot { \delta}_\alpha}}|$ and $\varphi_{k} = {\rm arg}\left( \sum_{\alpha}e^{i k \cdot \delta_\alpha}\right)$.
We will for simplicity now assume $\mu \sim t$ and focus on the lower $\pi$-band, but the same calculation is also valid for doping levels around the VHS at $\mu = -t$. By only keeping terms within the lower band, ignoring effects in the upper band along with any cross-terms, we arrive at: 
\begin{align}
\label{eq:H1band}
H'_z & = -h_z \sigma \sum_{k} a_{{ k}\sigma}^\dagger a_{{ k}\sigma} \nonumber \\
H'_\Delta & = - \sum_{{k},\alpha} \Delta_{\alpha} \cos({{k \cdot \delta}_\alpha}-\varphi_{k})(a_{{k}\uparrow}^\dagger a_{{-k}\downarrow}^\dagger) + {\rm H.c.}\nonumber \\
H'_\lambda & = \sum_{k\sigma \sigma'} \mathcal{L}_k\cdot {\bf s}_{\sigma \sigma'} a^\dagger_{k \sigma} a_{k \sigma'},
\end{align}
where $\mathcal{L}_k = \lambda_R{\rm Im}[e^{-i\varphi_k}(-\frac{\sqrt{3}}{2}e^{ik\delta_2} + \frac{\sqrt{3}}{2}e^{ik\delta_3},e^{ik\delta_1}-\frac{1}{2}e^{ik\delta_2}-\frac{1}{2}e^{ik\delta_3},0)]$.
The one-band Bogoliubov-de Gennes Hamiltonian $H'_{\rm ext} = H'_0 + H'_\Delta +H'_z + H'_\lambda$ can now be diagonalized and we find the eigenvalues
\begin{align}
\label{eq:Ek}
E(k) = \sqrt{(\mu-t\epsilon_k)^2 + \lambda_R^2 \mathcal{L}_k^2 + h_z^2 + |\Delta_k|^2 \pm 
2\sqrt{(\mu-t\epsilon_k)^2\lambda_R^2\mathcal{L}_k^2 + [(\mu-t\epsilon_k)^2 + |\Delta_k|^2]h_z^2}},
\end{align}
where $\Delta_k = \sum_{\alpha}\Delta_{\alpha} \cos({{k \cdot \delta}_\alpha}-\varphi_{k})$. Following the procedure in Ref.~\onlinecite{Sato10}, the zero energy values of Eq.~(\ref{eq:Ek}), or equivalently the bulk-gap closing condition, satisfy
\begin{align}
\label{eq:Egsupp}
(\mu - t|\epsilon_k|)^2 + \Delta_k^2 = h_z^2 + \lambda_R^2|{\mathcal L}_k|^2, \ \ \ |\Delta_k||\lambda_R \mathcal{L}_k| = 0,
\end{align}
which is the same as Eq.~(4) in the main text. From this equation we locate the only low-field bulk closing point to be at $(\mu - t)^2 + \Delta_k^2(M) = h_z^2$, where $\Delta_k(M) = 2\Delta$ for the $\Delta(1,e^{2\pi i /3},e^{4\pi i/3})$ order parameter. As seen in Fig.~4(b) in the main text this approximative bulk closing condition is accurate for small to moderately large $\lambda_R$.
%

\section{Disorder effects}
\label{sec:disorder}
Heavy doping of graphene will undoubtedly introduce some amount of disorder into the system. Disorder can affect the results derived in this Letter in several ways. First of all, sufficiently strong disorder will suppress the superconducting order parameter, this is especially true in non-$s$-wave superconductors. Secondly, disorder breaks the translational invariance and, thus, the two $d$-wave channels in graphene are no longer guaranteed to be degenerate. Related to this is the fact that there exists also an extended $s$-wave solution, which, in general, is heavily disfavored but in the presence of disorder might become more important. In addition to these bulk effects, disorder might also influence the edge properties of the \did\ state.  

In order to study the effect of disorder we model both the bulk and zigzag edges in the presence of Anderson disorder, i.e.~a locally fluctuating chemical potential:
\begin{align}
\label{eq:H0supp}
H_{0,{\rm dis}} & = -t  \sum_{\langle i,j\rangle,\sigma}  c^\dagger_{i\sigma}c_{j\sigma} + \sum_i (\mu + \delta\mu_i) c^\dagger_{i\sigma} c_{i\sigma}, 
\end{align}
where the local chemical potential variations $\delta\mu_i$ are distributed randomly within the interval $[-W,W]$, with $W$ being the disorder strength. This type of disorder model captures the effect of local charge inhomogeneities introduced by the doping. It is also reasonable to assume, as is done here, that  the disorder will in general be directionally independent, such that it does not single out one bond direction over the other.
We solve $H_{0,{\rm dis}}+H_\Delta$ for multiple disorder configurations in large bulk and edge samples and study how the superconducting order evolves with the disorder strength $W$. We fix $\mu = 1$ which maximizes the effect of the disorder, since both negative and positive deviations from the VHS causes the superconducting order parameter to decrease.
%
\begin{figure}[htb]
\includegraphics[scale = 0.9]{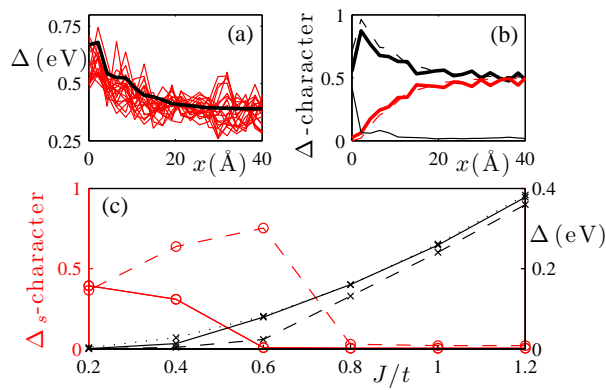}
\caption{\label{fig:dis} (Color online) (a) Order parameter $\Delta_1$ profiles at the zigzag edge for NN $J = 1.2t$ at the VHS for a 20 unit cell wide sample
with $W = 0.5t = 1$~eV disorder (red). Clean sample (thick black).
(b) Average character of the order parameter in (a): $d_1$ (thick black), $d_2$ (thick red), and $s$ (thin black). Dashed lines display the clean $d_1$ and $d_2$ results, respectively.
(c) Bulk $\Delta_1$ order parameter (right, black axis) as function of NN $J$ for clean (black dotted), $W = 0.2t = 0.5$~eV (black solid), and $W = 0.5t = 1$~eV (black dashed) and the corresponding $s$-wave character (left, red axis).
}
\end{figure}
In Fig.~\ref{fig:dis}(c) we plot on the right axis the superconducting order parameter $\Delta_1$ as function of the NN pairing potential $J$ for a $40 \times 40$~\AA\ bulk sample. The results are averaged over as many as 20 different disorder configurations. For $W = 0.2t = 0.5$~eV there is a suppression of the superconducting state for $J\leq 0.4t$, at which point the character of the superconducting state also changes from perfect \did\ to contain a significant amount of $s$-wave character (right axis). At $J = 0.4t$, $W$ is 18 times larger than $\Delta_1$ and thus the \did\ state survives disorder at least an order of a magnitude stronger than the superconducting gap. The appearance of a sizable $s$-wave component at very strong disorder is expected since isotropic states are more robust against disorder, but this also suppresses the overall superconducting order parameter.
For $W = 0.4t = 1$~eV we find the same scenario, with the \did\ state being suppressed into a weaker partial $s$-wave state at $J\leq 0.6t$, where $W$ is 12 times larger than $\Delta_1$.
Based on these results, we expect the \did\ state to survive essentially unchanged in the bulk in the presence of even moderately strong disorder.
In Figs.~\ref{fig:dis}(a,b) we plot the behavior at the edge for a representative $W = 1$~eV disorder configuration. Computational demands limit the size of the sample and we are forced to use a rather large $J = 1.2t$. Nonetheless, the disorder strength is still in this case almost 3 times larger than the bulk  $\Delta_1$ value. The $\Delta_1$ profile into the sample in Fig.~\ref{fig:dis}(a) shows a noticeable spatial variation, but still, the average is not suppressed much from the clean limit. As seen in Fig.~\ref{fig:dis}(b), the average character of the superconducting state is also essentially left unchanged by this relatively strong disorder.
We thus conclude that even moderately strong disorder does not influence the edge properties of the \did\ superconducting state in heavily doped graphene.


\end{document}